\documentclass{acm_proc_article-sp}

\usepackage[english]{babel}
\usepackage[latin1]{inputenc}
\usepackage{amsmath}
\usepackage{amsfonts}
\usepackage{amssymb}
\usepackage{graphicx}
\usepackage{subfigure}
\usepackage{moreverb}
\usepackage{multirow}
\usepackage[ruled,vlined]{algorithm2e}
\newcommand{\compresslist}{
  \vspace{-1em}
  \setlength{\itemsep}{3pt}
  \setlength{\parskip}{0pt}
  \setlength{\parsep}{0pt}
}

\begin{document}

\title{Evaluation of recommender systems in streaming environments}

\numberofauthors{3}
\author{
\alignauthor João Vinagre\\
	\affaddr{FCUP - Universidade do Porto}\\
	\affaddr{LIAAD - INESC TEC}\\
	\affaddr{Porto, Portugal}
	\email{jnsilva@inesctec.pt}
\alignauthor Alípio Mário Jorge\\ 
	\affaddr{FCUP - Universidade do Porto}\\
	\affaddr{LIAAD - INESC TEC}\\
	\affaddr{Porto, Portugal}
	\email{amjorge@fc.up.pt}
\alignauthor João Gama\\ 
	\affaddr{FEP - Universidade do Porto}\\
	\affaddr{LIAAD - INESC TEC}\\
	\affaddr{Porto, Portugal}
	\email{jgama@fep.up.pt}
}
\date{June 2014}

\conferenceinfo{}{Copyright is held by the author/owner(s). Workshop on 'Recommender Systems Evaluation: Dimensions and Design' (REDD 2014), held in conjunction with RecSys 2014. October 10, 2014, Silicon Valley, United States.}

\maketitle

\begin{abstract}
Evaluation of recommender systems is typically done with finite datasets. This means that conventional evaluation methodologies are only applicable in offline experiments, where data and models are stationary. However, in real world systems, user feedback is continuously generated, at unpredictable rates. Given this setting, one important issue is how to evaluate algorithms in such a streaming data environment. In this paper we propose a prequential evaluation protocol for recommender systems, suitable for streaming data environments, but also applicable in stationary settings. Using this protocol we are able to monitor the evolution of algorithms' accuracy over time. Furthermore, we are able to perform reliable comparative assessments of algorithms by computing significance tests over a sliding window. We argue that besides being suitable for streaming data, prequential evaluation allows the detection of phenomena that would otherwise remain unnoticed in the evaluation of both offline and online recommender systems.
\end{abstract}


\section{Introduction}
\label{sec:intro}

Usage-based algorithms for recommender systems rely on user-provided data. In a typical lab setting, this data has been previously collected from a system and consists of a finite set of user generated actions -- typically item ratings. These datasets contain enough data to objectively apply well studied methodologies to evaluate recommendation algorithms in a laboratory setting. However, it is increasingly consensual that the accuracy obtained by algorithms in controlled environments does not translate directly into good performance or overall user-perceived quality in a real-world production environment. 

We use the words ``online" and ``offline" to refer to the environment in which a system functions and/or is evaluated. \emph{Online} systems run in production mode in a real-world setting -- i.e. they provide or support an active online service to real users. \emph{Offline} systems essentially run in laboratorial or controlled conditions mainly for development and/or systematic evaluation purposes.

In this paper, we propose the application of a prequential evaluation methodology \cite{DBLP:conf/kdd/GamaSR09} for the evaluation of recommender systems. Prequential evaluation is primarily designed to evaluate algorithms that learn from continuous flows of data -- data streams. If we look at user-generated data typically used by recommendation systems, we can safely state that this user feedback is generated online at unpredictable rates and ordering, and is potentially unbounded. This is the exact definition of a data stream \cite{DBLP:conf/dmkd/DomingosH01}. On the one hand, this motivates the use of incremental algorithms, since batch rebuilding of predictive models may eventually become too expensive. On the other hand it also motivates a reflection on the applicability of classic evaluation methodologies to non-stationary models. This is because we are no longer trying to evaluate one model within a well delimited time frame, but rather a continuous and ever unfinished learning process. Prequential evaluation is especially adequate for this kind of setting. Nevertheless it is also applicable offline, with static datasets, as illustrated in this paper (Sec. \ref{sec:exp}).

Holdout methods are widely used in the evaluation of recommender systems, however they are designed for batch algorithms and are not directly applicable in a non-stationary setting. Indeed, if data points are constantly being generated we can only take subsets of the available data and evaluate the algorithms on those subsets. Moreover, if we decide to implement an incremental algorithm, some issues on its evaluation, such as dataset ordering or recommendation bias, are not easy to circumvent (see Sec. \ref{sec:classic_eval}).

Prequential evaluation does not require data pre-processing and is not restrictive in terms of evaluation criteria. It may include accuracy metrics typically used offline -- e.g. precision/recall, RMSE --, but also allows online measurements of complex user interaction behavior or acceptance feedback, thereby including actual users in the evaluation process. By collecting diverse statistics, it is also possible to combine several important dimensions of the evaluation of recommender systems, such as novelty, serendipity, diversity, trust and coverage \cite{DBLP:reference/rsh/ShaniG11}, or to collect online A/B testing feedback data \cite{DBLP:journals/datamine/KohaviLSH09}. We illustrate the use of prequential evaluation by observing a simple accuracy metric over time for the comparison of three algorithms, along with pairwise statistical significance tests. 

To our knowledge, prequential evaluation has only been used for recommendation algorithms very recently in our work \cite{DBLP:conf/um/VinagreJG14} and in \cite{DBLP:conf/wims/SiddiquiTSSM14}. The first essentially uses the evaluation process described in this paper, and is a more direct application of the prequential methodology used in data stream mining. The second proposes a hybrid method that uses both holdouts and prequential evaluation in mini-batches. While in \cite{DBLP:conf/um/VinagreJG14} our focus is essentially on the proposed algorithm, this paper focuses on the evaluation methodology itself with more detail, with the intent to raise discussion on the evaluation issues of incremental algorithms. We also illustrate the applicability of statistical significance tests to compare algorithms over time.

The remainder of this paper is structured as follows. In Sec. \ref{sec:classic_eval} we describe the traditional batch evaluation methodologies. Prequential evaluation is described in Sec. \ref{sec:prequential_eval}. We present an illustrative evaluation of three incremental recommendation algorithms in Sec. \ref{sec:exp}. Finally we conclude in Sec. \ref{sec:conclusions}.

\section{Evaluation methodologies}
\label{sec:classic_eval}

Traditionally, holdout methods are used in the batch evaluation of recommender systems. They begin by splitting the ratings dataset in two subsets -- training set and testing set -- randomly choosing data elements from the initial dataset. The training set is initially fed to the recommender algorithm to build a predictive model. 

There is some variety of offline protocols to evaluate accuracy, however they are essentially variations of holdout strategies. Generally, these protocols ``hide'' a subset of ratings given by each user in the test set. These hidden interactions form a hidden set. Algorithms are evaluated by measuring the difference between predictions and the actual observations in the hidden set.

\subsection{Issues with batch evaluation} 
\label{subsec:issues_eval}

Given the described offline evaluation methodology we identify the following issues:

\begin{itemize}
  \item \emph{Dataset ordering}: randomly selecting data for training and test, as well as random hidden set selection, shuffles the natural sequence of the data. Algorithms designed to deal with naturally ordered data cannot be rigorously evaluated if datasets are shuffled. One straightforward solution is simply not to shuffle data. That is, to pick a moment in time or a number of ratings in the dataset as the split point. All ratings given before the split point are used to train the model and all subsequent ratings are used as testing data. One awkwardness with this approach is how to select the hidden set. In \cite{DBLP:reference/rsh/ShaniG11} and \cite{Lathia2010} the authors suggest that all ratings in the test set should be hidden;
  \item \emph{Time awareness}: shuffling data potentially breaks the logic of time-aware algorithms. For example, by using future ratings to predict past ratings. This issue may as well be solved by keeping the chronological order of data;
  \item \emph{Incremental updates}: incremental algorithms perform incremental updates of their models as new data points become available. This means that neither models or training and test data are static. Models are continuously being readjusted with new data. As far as we know to this date, the only contributions in the field of recommender systems that explicitly address this issue are \cite{DBLP:conf/um/VinagreJG14} and \cite{DBLP:conf/wims/SiddiquiTSSM14}. This issue has already been addressed in the field of data stream mining \cite{DBLP:conf/kdd/GamaSR09,DBLP:journals/ml/GamaSR13};
  \item \emph{Session grouping}: most natural datasets, given their unpredictable ordering, require some pre-processing to group ratings either by user or user session in order to use offline protocols. As data points accumulate, it eventually may become too expensive to re-group them. This is true also for any other kind of data pre-processing task;
  \item \emph{Recommendation bias}: in online production systems, user behavior is -- at least expectedly -- influenced by recommendations themselves. It is reasonable to assume, for instance, that recommended items will be more likely followed than if they were not recommended. Simulating this offline usually requires complicated user behavior modeling which can be expensive and prone to systematic error. One way to evaluate the actual impact of a recommender system is to conduct user surveys and/or A/B testing \cite{DBLP:journals/datamine/KohaviLSH09,DBLP:journals/ijmir/DominguesGJLVLS13,DBLP:journals/umuai/KnijnenburgWGSN12,DBLP:journals/umuai/PuCH12}. 
\end{itemize}

The above limitations, along with other known issues \cite{DBLP:reference/rsh/ShaniG11,DBLP:conf/chi/McNeeRK06,DBLP:journals/tois/HerlockerKTR04}, weaken the assumption that user behavior can be accurately modeled or reproduced in offline experiments. From a business logic perspective \cite{DBLP:conf/iccsa/FelixSJV14} offline evaluation may also not be timely enough to support decision making. These issues motivate the research of alternative or complementary evaluation methodologies.

\subsection{Offline evaluation protocols and metrics}
\label{subsec:offline_protocols}

One important consideration about the evaluation of a recommendation algorithm is the type of problem or task being approached. When dealing with explicit numeric ratings, the first task of the algorithm is to accurately predict unknown ratings. This is usually referred to as a \emph{rating prediction} task and is most naturally seen as a regression problem. One way to assess the accuracy of rating prediction algorithms is to measure the error of predicted ratings, given the true values in the hidden set \cite{DBLP:conf/chi/ShardanandM95}, using metrics such as Mean Absolute Error (MAE) and Root Mean Squared Error (RMSE). This protocol is in fact the most common approach, having been used in highly popularized competitions such as the Netflix prize \cite{Bennett07thenetflix} and KDD-Cup 2011 \cite{DBLP:journals/jmlr/DrorKKW12}.

However, numeric ratings may not be available. In such cases, the data usually consists of a record of positive-only user-item interactions. The task is then to predict good items to recommend. This problem is usually referred to as \emph{item prediction}. Item prediction problems can be evaluated both as classification and ranking problems. Accuracy is measured by matching recommendation lists to the true hidden items for each user. Typically, classification metrics such as Precision, Recall and F-measure or ranking metrics such as Mean Average Precision (MAP) \cite {DBLP:conf/www/McFeeBEL12} or Normalized Discounted Cumulative Gain (NDCG) \cite{DBLP:conf/nips/WeimerKLS07} are used. The first protocols used for the evaluation of item recommendation problems are the ones known as \emph{All-but-N} and \emph{Given-N} \cite{DBLP:conf/uai/BreeseHK98}. The \emph{All-but-N} protocol hides exactly $N$ items from each user in the test set. One popular sub-protocol is the \emph{All-but-One} protocol, which hides exactly one item from each user in the test set. The \emph{Given-N} protocol keeps exactly $N$ items in the test set and hides all others.

\section{Prequential evaluation}
\label{sec:prequential_eval}

Given the problems listed in Section \ref{subsec:issues_eval}, we propose a prequential approach \cite{DBLP:conf/kdd/GamaSR09}, especially suited for the evaluation of algorithms in a non-stationary environment. Essentially, the prequential method consists of a \emph{test-then-learn} procedure that runs for each new data point. Given a newly observed data point, a prediction is made and tested -- e.g. measuring error. Then, the data point is used to update the model. 

In this paper, we illustrate prequential evaluation with item prediction recommenders. The item prediction task consists of selecting good items for recommendation, which are typically presented to the user as a ranked list. In this task prequential evaluation consists of the folowing steps:

\begin{enumerate}
  \compresslist
  \item If $u$ is a known user, use the current model to recommend $N$ items to $u$, otherwise go to step 3;
  \item Score the recommendation list given the true observed item $i$;
  \item Update the model with the observed event;
  \item Proceed to the next event in the dataset;
\end{enumerate} 

In its strict formulation, prequential evaluation does not require step 3. Indeed, one may not wish to update the model at every single observation, or  ever. This allows the comparison between different types of algorithms, for example, incremental vs. batch algorithms.

This protocol provides several benefits:
\begin{itemize}
  \compresslist
  \item It allows continuous monitoring of the system's performance over time;
  \item Several metrics can be captured simultaneously;
  \item If available, user feedback can be included in the loop;
  \item Real-time statistics can be integrated in the algorithms' logic -- e.g. automatic parameter adjustment, drift/shift detection, triggering batch retraining;
  \item In ensembles, relative weights of individual algorithms can be adjusted;
  \item The protocol is applicable to both item prediction and rating prediction;
  \item By being applicable both online and offline, experiments are naturally reproducible if the same data sequence is available.
\end{itemize}

In an offline experimental setting, an overall average of individual scores can be computed at the end -- because lab datasets are inevitably finite -- and on different time horizons. For a recommender running in a production system, this process allows us to follow the evolution of the recommender by keeping online statistics of any number of chosen metrics. Thereby it is possible to depict how the algorithm's performance evolves over time. In Sec. \ref{sec:exp} we present both the overall average score and complement it with plots of the evolving score using a simple moving average of an accuracy metric.

One challenging aspect of this method is that it only evaluates over a single item at each step, potentially failing to recognize other possible good recommendations. If item $i$ is not recommended at the time the observation is made, the score will naturally be 0. However, other items within the $N$ recommendations may occur in future observations for that user. In other words, the protocol exclusively evaluates how well the model predicts the next observation, ignoring all subsequent ones. Although this is a somewhat challengingly strict protocol, we have performed experiments by matching the recommended items with not just the current, but all future observations for each user -- only possible offline --, and found that overall scores do not improve substantially. However, this strictness of the protocol may potentially have a higher impact with other metrics or data. One way to relax this, is to match the active observation not only with the current prediction, but also with a set of previous predictions. One other possible approach is to use a hybrid evaluation method such as in \cite{DBLP:conf/wims/SiddiquiTSSM14}.

\section{Applying the methodology}
\label{sec:exp}

To illustrate the usefulness of prequential evaluation in recommender systems, we perform a set of experiments using this protocol. We use three item recommendation algorithms that learn recommendation models incrementally as user feedback data becomes available. These algorithms are designed to process positive-only feedback -- also known as binary feedback. However, we emphasize that this is not a restriction of the evaluation protocol, since it is possible to use the exact same methodology for rating prediction problems as well.

\subsection{Datasets}
\label{subsec:datasets}

We use four distinct datasets, described in Table \ref{tab:datasets}. All datasets consist of a chronologically ordered set of pairs in the form $<user, item>$. Music-listen and Lastfm-600k consist of music listening events obtained from two distinct sources, where each tuple corresponds to a music track being played by a user. Music-playlist consists of a timestamped log of music track additions to personal playlists. MovieLens-1M is well known dataset\footnote{http://www.grouplens.org, 2003} consisting of timestamped movie ratings in a 1 to 5 rating scale. To use this dataset in an item prediction setting, since we intend to retain only positive feedback, movie ratings below the maximum rating 5 are excluded. Lastfm-600k consists of the first 8 months of activity observed in the Last.fm\footnote{http://last.fm} dataset originally used in \cite{DBLP:books/daglib/0025137}. Both Music-listen and Music-playlist are extracted from the Palco Principal\footnote{http://www.palcoprincipal.com} website, a social network dedicated to non-mainstream music enthusiasts and artists.

\begin{table}
\centering
\footnotesize
\begin{tabular*}{0.45\textwidth}{@{\extracolsep{\fill}} l r r r r r }
\hline
Dataset & Events & Users & Items & Sparsity \\
\hline
Music-listen & 335.731 & 4.768 & 15.323 & 99,90\% \\
\hline
Lastfm-600k & 493.063 & 164 & 65.013 & 99,11\% \\
\hline
Music-playlist & 111.942 & 10.392 & 26.117 & 99,96\% \\
\hline
MovieLens-1M & 226.310 & 6.014 & 3.232 & 98,84\% \\
\hline
\end{tabular*}
\caption{Dataset description}\normalsize
\label{tab:datasets}
\end{table}

\begin{figure*}[ht!]
\centering
\includegraphics[width=16cm,height=12cm]{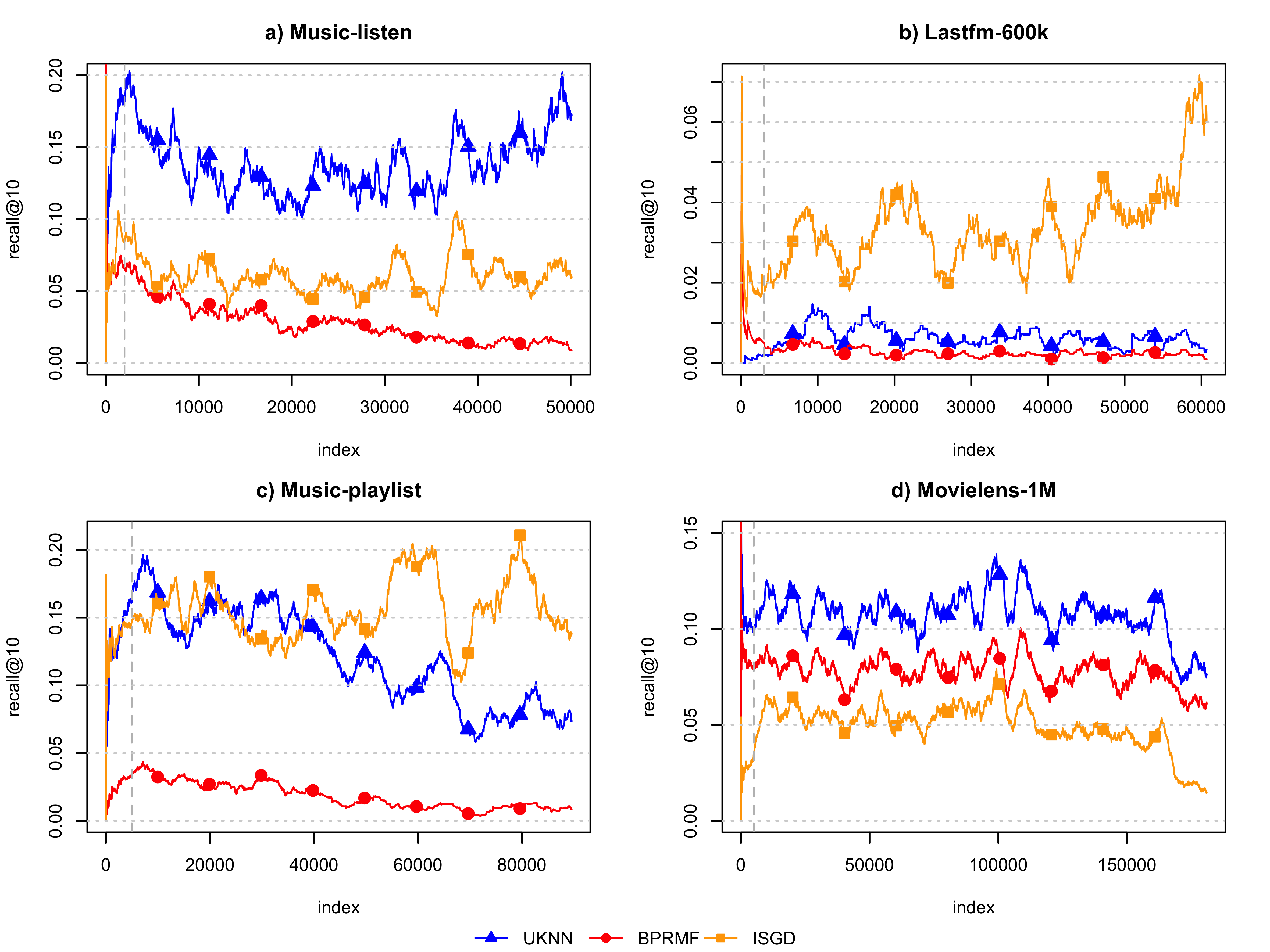}
\caption{Evolution of recall@10 with four datasets. The plotted lines correspond to a moving average of the recall@10 obtained for each prediction. The window size $n$ of the moving average is a) $n=2000$, b) $n=3000$, c) $n=5000$ and d) $n=5000$. The first $n$ points are delimited by the vertical dashed line and are plotted using the accumulated average. Plots a) and b) do not include repeated events in the datasets.}
\label{fig:results_evolving}
\end{figure*}

\begin{figure*}[ht!]
\centering
\includegraphics[width=16cm,height=8cm]{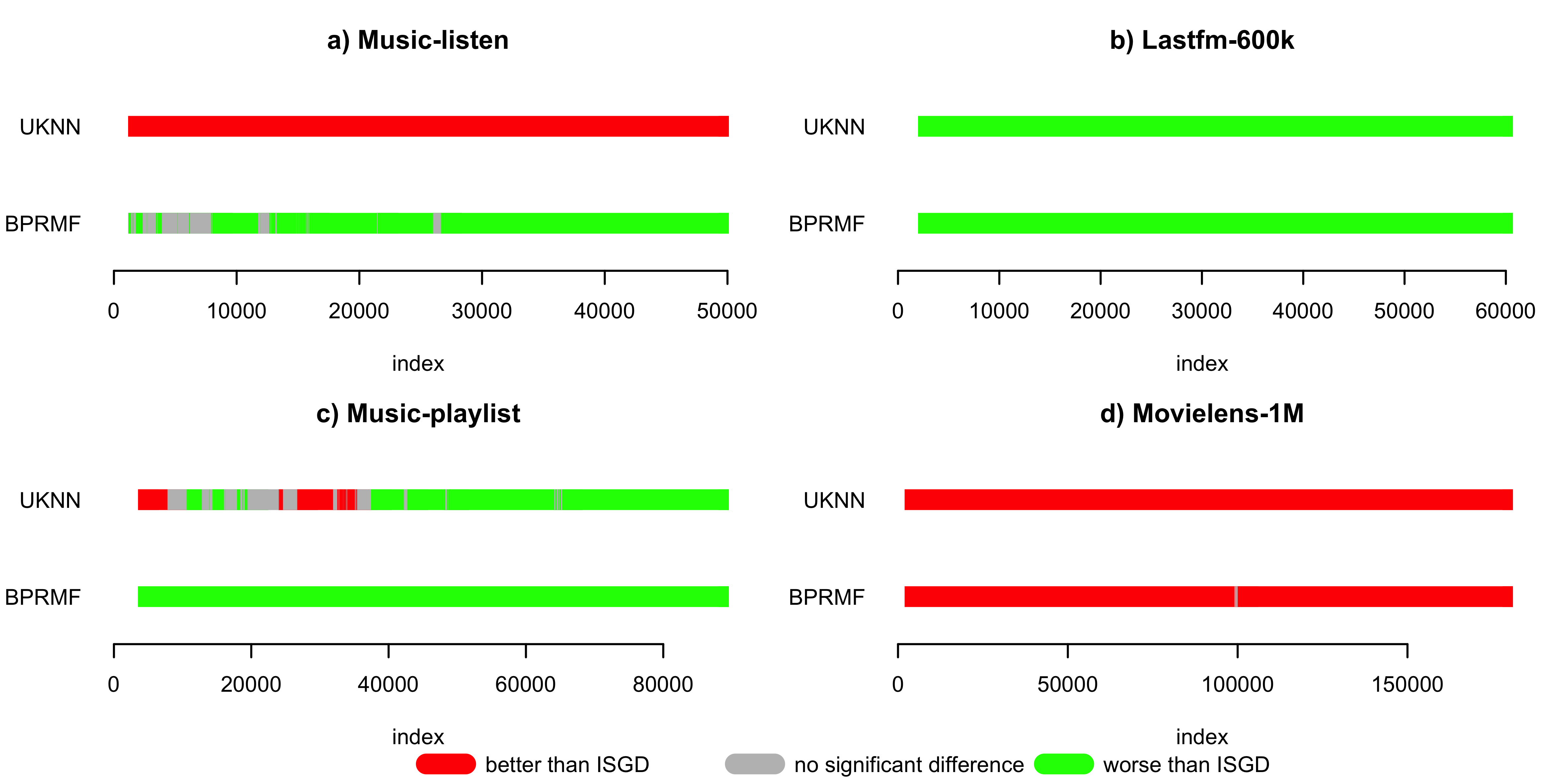}
\caption{Signed McNemar test of ISGD against BPRF and UKNN. Test is computed over a sliding window of $n$ observations with a) $n=2000$, b) $n=3000$, c) $n=5000$ and d) $n=5000$.}
\label{fig:mcnemar}
\end{figure*}

\subsection{Algorithms used and overall accuracy}
\label{subsec:overall_res}

Using the prequential approach described in Sec. \ref{sec:prequential_eval}, we compare the online accuracy of 3 incremental item recommendation algorithms: ISGD \cite{DBLP:conf/um/VinagreJG14}, BPRMF \cite{DBLP:conf/uai/RendleFGS09} and a classic incremental user-based neighborhood algorithm \cite{DBLP:conf/webi/MirandaJ08}, all implemented in MyMediaLite\footnote{http://mymedialite.net} \cite{DBLP:conf/recsys/GantnerRFS11}. We measured accuracy with recall at cut-off 10 -- denoted by recall@10.

\begin{table}\footnotesize
\begin{center}
\begin{tabular}{l|l|r|r}
\hline
Dataset & Algorithm & Recall@10 & Update time \\
\hline
	\multirow{4}{*}[3pt]{Music-listen}
	& BPRMF & 0.028 & 0.846 ms \\
	& ISGD & 0.061 & \textbf{0.118 ms} \\
	& UserKNN & \textbf{0.139} & 328.917 ms \\
\hline
	\multirow{4}{*}[3pt]{Lastfm-600k}
	& BPRMF & 0.003 & 28.061 ms \\
	& ISGD & \textbf{0,034} & \textbf{1.106} ms \\
	& UserKNN & 0.006 & 290.133 ms \\
\hline
	\multirow{4}{*}[3pt]{Music-playlist}
	& BPRMF & 0.020 & 1.889 ms \\
	& ISGD & \textbf{0.171} & \textbf{0.949 ms} \\
	& UserKNN & 0.132 & 190.250 ms \\
\hline
	\multirow{4}{*}[3pt]{Movielens-1M}
	& BPRMF & 0.080 & 0.173 ms \\
	& ISGD & 0.050 & \textbf{0.016 ms} \\
	& UserKNN & \textbf{0.110} & 84.927 ms \\
\hline
\end{tabular}\normalsize
\caption{Overall results. Best performing algorithms are highlighted in bold for each dataset. Update times are the average value of the update time for all data points.}
\label{tab:results_overall}
\end{center}
\end{table}


Overall results with average update times are presented in Tab. \ref{tab:results_overall}. These are possible to obtain in offline experiments, given that lab datasets are finite. However, in online production systems these results can only be interpreted as a snapshot of the algorithms' performance within a predefined time frame. 

\subsection{Accuracy over time}
\label{subsec:monitoring}

One valuable feature of our adopted evaluation protocol is that it allows the monitoring of the learning process as it evolves over time. To do that, we need to maintain statistics of the outcome of the predictions. We study how the algorithms' accuracy evolves over time by depicting in Fig. \ref{fig:results_evolving} a moving average of the recall@10 metric. The moving average sizes are chosen to obtain clear lines in Fig. \ref{fig:results_evolving}, for illustrative purposes. We do not argue that these values are any better than others.

The plotted evolution of the algorithms with each dataset generally confirms overall results, however more details become available. For instance, although the overall averages of ISGD and UKNN are relatively close with the Music-playlist dataset, Fig. \ref{fig:results_evolving} c) shows that these algorithms behave quite differently, starting with a very similar accuracy level and then diverging substantially. Although this kind of observation could be important for a rigorous evaluation, it is diluted in a single average in Table \ref{tab:results_overall}.

\subsection{Statistical significance over time}
\label{subsec:mcnemar}

We also depict in Fig. \ref{fig:mcnemar} statistical significance tests using the signed McNemar test over sliding windows \cite{DBLP:conf/kdd/GamaSR09} of the same size as the ones used for the moving averages used in Fig. \ref{fig:results_evolving}. We set a significance level of 1\%. Because McNemar is a pairwise test, a complete comparative assessment with four datasets and three algorithms would require 12 tests. However, to avoid multiple tests we can compare one proposed algorithm with existing ones. Alternatively we can use p-value corrections. In this illustrative experiment we compare the ISGD algorithm with the other two on the four datasets, which yields 8 tests. The main observation from Fig. \ref{fig:mcnemar} is that the most of the apparent diferences in Fig. \ref{fig:results_evolving} are statistically significant. However, the visualization of the McNemar test clarifies some comparisons.

The online monitoring of the learning process allows a more detailed evaluation of the algorithms' performance. Figure \ref{fig:results_evolving} reveals phenomena that would otherwise be hidden in a typical batch evaluation. We consider that this finer grained evaluation process provides a deeper insight into the learning processes of predictive models.

\section{Conclusions}
\label{sec:conclusions}

In this paper, we propose a prequential evaluation framework to monitor evaluation metrics of recommender systems as they continuously learn from a data stream. To illustrate its applicability and appropriateness we use this framework to compare three incremental recommendation algorithms. We notice that our evaluation method allows a finer grained assessment of algorithms, by being able to continuously monitor the outcome of the learning process. Moreover, it is possible to integrate multiple measures simultaneously in the evaluation process, thereby evaluating multiple dimensions. We also show the applicability of statistical significance tests.  

\section{Acknowledgements}
This work is partially funded by FCT/MEC through 
PIDDAC and ERDF/ON2 within project NORTE - 07 - 0124 - FEDER - 000059 and through the COMPETE Programme (operational programme for competitiveness) and by National Funds through the FCT - Funda\c{c}\~ao para a Ci\^encia e a Tecnologia (Portuguese Foundation for Science and Technology) within project FCOMP - 01 - 0124 - FEDER - 037281. The first author's work is funded by the FCT grant SFRH / BD / 77573 / 2011. The authors wish to thank Ubbin Labs, Lda. for kindly providing data from Palco Principal.

\bibliographystyle{abbrv}
\bibliography{refs}

\end{document}